\documentclass[aps,prl,twocolumn,showpacs,superscriptaddress]{revtex4}
\usepackage{times}
\usepackage{epsfig}

\begin{document}

\title{Quantum Limits of Stochastic Cooling of a Bosonic Gas}

\author{D. Ivanov}

\altaffiliation{Also at: Russian Center of Laser Physics,
  St.-Petersburg State University, 5 Ulianovskaya, Pedrodvoretz, St.
  Petersburg, Russia}

\affiliation{Emmy--Noether Nachwuchsgruppe ``Kollektive Quantenmessung
  und R\"uckkopplung an Atomen und Molek\"ulen'', Fachbereich Physik,
  Universit\"at Rostock, Universit\"atsplatz 3, D-18051 Rostock,
  Germany}

\author{S. Wallentowitz} 

\email{sascha.wallentowitz@physik.uni-rostock.de}

\affiliation{Emmy--Noether Nachwuchsgruppe ``Kollektive Quantenmessung
  und R\"uckkopplung an Atomen und Molek\"ulen'', Fachbereich Physik,
  Universit\"at Rostock, Universit\"atsplatz 3, D-18051 Rostock,
  Germany}

\author{I.A. Walmsley}

\affiliation{Clarendon Laboratory, University of Oxford, Parks Road,
  Oxford OX1 3PU, UK}

\date{\today}

\begin{abstract}
  The quantum limits of stochastic cooling of trapped atoms are
  studied. The energy subtraction due to the applied feedback is shown
  to contain an additional noise term due to atom-number fluctuations
  in the feedback region. This novel effect is shown to dominate the
  cooling efficiency near the condensation point.  Furthermore, we
  show first results that indicate that Bose--Einstein condensation
  could be reached via stochastic cooling.
\end{abstract}


\maketitle

Ultracold atomic gases are generated by laser
cooling~\cite{laser-cooling}, evaporative cooling~\cite{evap-cooling}
or sympathetic cooling~\cite{symp-cooling}. These techniques have been
successful in preparing Bose--Einstein condensed~\cite{bec} and Fermi
degenerate~\cite{fermi} atomic gases. These states of matter have
proven important in both the fundamental physics of weakly interacting
many-body systems~\cite{mott} and in
applications~\cite{atom-laser,bec-micro}.  Nonetheless, these cooling
methods have some limitations. For example, laser cooling requires
closed-cycle transitions, evaporative cooling leads to a loss of a
significant fraction of the atoms, and sympathetic cooling requires
careful selection of buffer species with sufficiently large scattering
cross sections.

A much more general strategy that avoids these limitations is
stochastic cooling~\cite{stochastic-cooling}. It is a Maxwell-demon
strategy that uses information obtained from measurement to coherently
reduce the energy of part of the system.  Stochastic cooling is based
on the repeated application of a feedback loop. The cooling is
obtained by the combination of measurement and controlled Hamiltonian
interaction during the feedback operation. The interaction provides an
energy exchange with an external field and the preceeding measurement
ensures the irreversibility of this exchange. In this way the feedback
mechanism may be thought of as acting as a dissipative reservoir.
Classically cooling occurs due to the extraction of information on the
phase-space localization of particles, and its subsequent use to
reposition the particles, leading to a phase-space
compression~\cite{stochastic-cooling}.

In high-energy physics it has been employed for cooling the transverse
degree of freedom of a particle beam~\cite{stochastic-cooling}. A
measurement of the transverse momentum of a fraction of the particles
is made.  A control field sets the momentum to zero, which together
with a subsequent remixing of the particles leads to phase-space
compression and cooling of the transverse motion.  Recently,
stochastic cooling was proposed for trapped atoms and it was shown
that both momentum measurement and shift could be realized by optical
fields~\cite{stochastic-raizen}.  The required remixing of atoms is
provided here by the oscillation in the trap. Moreover, interactions
between atoms, such as collisions, may provide a further enhancement
of this remixing. Classical calculations for a 1D atomic gas showed a
pronounced cooling effect~\cite{stochastic-raizen}, so that stochastic
cooling may perhaps be an alternative to standard cooling methods.

However, to best of our knowledge, it is not yet known to what
temperatures such a method eventually will cool the atoms and whether,
for example, Bose--Einstein condensation can be reached.  Technical
heating effects, that are inherent in the proposed optical
implementation, have been discussed~\cite{stochastic-raizen}.
However, the fundamental limits of such a cooling method, due to the
discreteness of atoms and their quantum correlations, seem to be
unexplored. A reason for the lack of knowledge of these limits may be
that many-atom correlations play a central role. The atoms cannot be
treated as individual entities, which would considerably simplify a
theoretical description, but require a quantum many-atom
description~\cite{wal-feedback}.

Measurement in quantum mechanics always leads to a back-action in the
conjugate variable, and one might expect that this will saturate the
cooling at ultralow temperatures.  It will be shown in this Letter
that a further fundamental heating mechanism arises from the quantum
fluctuations of the number of atoms in the feedback region.  To reach
ultralow temperatures, this heating has to be circumvented, which will
be shown to be possible by the choice of the feedback region. We treat
stochastic cooling in a fully quantum-field theoretical way and show
initial results that indicate the possibility of reaching
Bose--Einstein condensation.

\begin{figure}
  \begin{center}
    \includegraphics[width=0.3\textwidth]{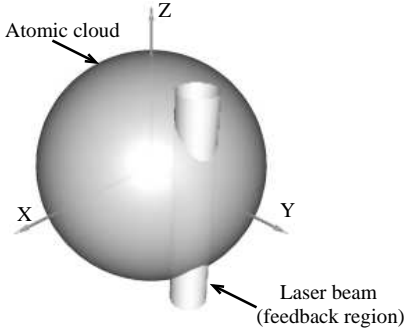}    
    \caption{Geometry of the feedback setup. The laser beam is aligned
      along the $z$ axis, it determines the size of the feedback
      region.}
    \label{fig:geometry}
  \end{center}
\end{figure}

The feedback loop of stochastic cooling consists of a measurement of
the total momentum $\hat{P}_w$ of the atoms in the feedback region and
a subsequent interaction in the same spatial region that compensates
for the observed value $P$, i.e., shifts the total momentum to zero.
The spatial region where atoms are subject to the feedback is defined
by the beam of a laser field that implements the measurement and
subsequent shift, cf.~Ref.~\cite{stochastic-raizen}.  For simplicity
we consider the case where the laser beam is aligned along the $z$
axis and the beam waist $w$ shall be a step-like function in the
$xy$-plane, cf.~Fig.~\ref{fig:geometry}.
Note that in $z$-direction no spatial restriction is assumed.

The measurement and subsequent shift shall occur without time delay
and on timescales much smaller than the intrinsic timescale of the
system $1/\omega_0$, where $\omega_0$ is the trap frequency of the 3D
isotropic, harmonic potential. Then the entire feedback loop
approximately acts instantaneously in time. Furthermore, we take into
account the resolution $\sigma$ of the measurement of momentum
$\hat{P}_w$, that is determined by external optical fields
implementing the measurement.

The feedback acts on the many-atom density operator $\hat{\varrho}$ as
\begin{equation}
  \label{eq:fb-mp}
  \hat{\varrho}_+ = \int \! dP \, \hat{U}(P) \, \hat{M}(P \!-\!
  \hat{P}_w) \,
  \hat{\varrho}_- \, \hat{M}^\dagger(P \!-\! \hat{P}_w) 
  \, \hat{U}^\dagger(P) ,
\end{equation}
where $\hat{\varrho}_\mp$ are the many-atom density operators before
($-$) and after ($+$) the feedback. The measurement is described by
the positive operator-valued measure $\hat{M}$, where $|M(P)|^2$ is a
Gaussian of width $\sigma$. The unitary operator $\hat{U}(P) \!=\!
\exp[ i P \hat{Q}_w]$ implements the subsequent shift of the measured
momentum $P$ to zero~\cite{remark}. The final density operator is
found by averaging over all possible measurement outcomes.

Using the bosonic atom-field operator $\hat{\phi}({\mathbf r})$ with
commutator $[\hat{\phi}({\mathbf r}), \hat{\phi}^\dagger({\mathbf
  r}')] \!=\! \delta^{(3)}({\mathbf r} \!-\!  {\mathbf r}')$, the
measured total momentum $\hat{P}_w$ and the center-of-mass $\hat{Q}_w$
of atoms in the feedback region are
\begin{eqnarray}
  \label{eq:Pw}
  \hat{P}_w & = & \int_w \!\! dA \int \! dz \, 
  \hat{\phi}^\dagger({\mathbf r}) 
  \, (-i\partial_z) \, \hat{\phi}({\mathbf r}) ,
  \\ 
  \label{eq:Qw}
  \hat{Q}_w & = & \frac{1}{N_{\rm e}} 
  \int_w \!\! dA \int \! dz \,
  \hat{\phi}^\dagger({\mathbf r}) \, z \, 
  \hat{\phi}({\mathbf r}) ,
\end{eqnarray}
where $dA$ is integrated over the beam waist $w$.  Their commutator is
$[\hat{Q}_w, \hat{P}_w] \!=\! i \hat{N}_w / N_{\rm e}$, where the
operator of the number of atoms in the feedback region is
\begin{equation}
  \label{eq:Nw}
  \hat{N}_w = \int_w \!\! dA \int \! dz \, 
  \hat{\phi}^\dagger({\mathbf r}) \, \hat{\phi}({\mathbf r}) .
\end{equation}
Note that in Eq.~(\ref{eq:Qw}) the estimated number of atoms $N_{\rm
  e}$ is used and not the proper operator $\hat{N}_w$: Since, roughly
speaking, in the Schr\"odinger picture $\hat{U}(P) \!=\! \prod_i
\exp(i P/N_{\rm e} \hat{q}_i )$, where $\hat{q}_i$ is the coordinate
of the $i$th atom, each atom is shifted by $-P/N_{\rm e}$. That is,
since the true atom number is unknown --- it is not measured --- an
estimate for the atom number, $N_{\rm e}$, has to be employed for
properly shifting each atom's momentum.

For cooling we are interested in the average energy that is subtracted
by the feedback from the set of non-interacting atoms,
\begin{equation}
  \label{eq:dE-def0}
  \Delta E = \langle \hat{H} \rangle_+ - \langle \hat{H}
  \rangle_-, 
\end{equation}
Here $\hat{H}$ is the Hamiltonian of the atoms in the harmonic trap
potential and $\langle \ldots \rangle_\pm \!=\! {\rm Tr}(
\hat{\varrho}_\pm \ldots )$.  After a detailed calculation, using the
results of Ref.~\cite{wal-feedback}, we find that $\Delta E$ consists
of both cooling and heating terms.

Firstly, consider an expansion of the kinetic energy of the center of
mass of the atoms in the feedback region, around the estimated atom
number $N_{\rm e}$: 
\begin{equation}
  \label{eq:expansion}
  \hat{P}_w^2 / (2m \hat{N}_w) =  \hat{P}_w^2 /
  (2m N_{\rm e}) ( 1 \!-\!  \delta \hat{N}_w / N_{\rm e} \!+\! \ldots ) , 
\end{equation}
where $\delta \hat{N}_w \!=\!  \hat{N}_w \!-\! N_{\rm e}$.  Following
Ref.~\cite{wal-feedback} we find that only the zeroth- and first-order
terms of this expansion appear in the energy change $\Delta E$: the
expectation values of higher-order corrections exactly cancel each
other.  The leading term represents the cooling of the system by
feedback. For this term, the energy removed is $- \langle \hat{P}_w^2
\rangle_- / (2m N_{\rm e})$. With $m$ being the atomic mass, $N_{\rm
  e} m$ is the estimated total mass of the atoms in the feedback
region, and $\hat{P}_w$ is their total momentum.  Thus this is the
negative (estimated) kinetic energy of the center-of-mass of the atoms
in the feedback region. 
According to~(\ref{eq:Pw}), this term contains 
atom-atom correlations of the form $\langle \hat{\phi}^\dagger({\bf
  r}_1) \hat{\phi}({\bf r}_2) \hat{\phi}^\dagger({\bf r}_3)
\hat{\phi}({\bf r}_4) \rangle$ --- a clear indication that in the
quantum regime stochastic cooling cannot be described as a single-atom
problem.

The first-order correction in Eq.~(\ref{eq:expansion}) is found to
give rise to a heating contribution to $\Delta E$ of the form $\langle
\delta \hat{N}_w \hat{P}_w^2 \rangle_- / (2m N_{\rm e}^2)$. From
Eq.~(\ref{eq:expansion}) it can be seen, that this heating is due to
non-optimal shifts of total momentum produced by atom-number
fluctuations around the estimated value $N_{\rm e}$. Another way to
see this is to consider a many-atom quantum state $|\Psi\rangle$ after
a perfect measurement ($\sigma \!=\! 0$) of $\hat{P}_w$ with outcome
$P$. It shall also be an eigenstate of $\hat{N}_w$ with $N$ atoms in
the feedback region, i.e., $|\Psi\rangle \!=\!  |P,N\rangle$.  After
the momentum shift the state becomes $\hat{U}(P) \, |P,N\rangle \!=\!
| P ( 1 \!-\! N/N_{\rm e} ) , N \rangle$.  That is, the momentum will
be shifted to zero only if $N_{\rm e} \!=\!  N$.  Given that the
system in general is in a state of imprecise atom number, choosing
$N_{\rm e} \!=\!  \langle \hat{N}_w \rangle$ will only produce the
correct momentum shift on average.  Therefore, atom-number
fluctuations in the feedback region are transferred into momentum
fluctuations, that lead to the observed residual kinetic energy. This
poses a fundamental limit to the perfect operation of the feedback
loop and thus to stochastic cooling~\footnote{Additional well-resolved
  measurements of $\hat{N}_w$ in each feedback could avoid this, if
  done without additional heating.}.

Moreover, since a measurement is involved in the control loop, $\Delta
E$ contains also measurement-induced heating terms: Since the total
momentum is measured with resolution $\sigma$, this fluctuations of
the total momentum leads to a residual kinetic energy of the center of
mass of the atoms in the feedback region. Together with the correction
factor $\langle \hat{N}_w \rangle_- / N_{\rm e}$ that accounts for the
difference between estimated and average atom number, the resulting
heating term is $\sigma^2 / (2mN_{\rm e}) \!\times\! \langle \hat{N}_w
\rangle_- / N_{\rm e}$. Clearly a measurement of total momentum with
resolution $\sigma$ induces back-action noise $\propto \sigma^{-1}$
into the center of mass of the measured atoms. This noise leads to an
increase in potential energy (i.e. heating) of the center of mass in
the harmonic trap, which is found to be $m N_{\rm e} \omega_0^2 / (8
\sigma^2) \!\times\! \langle \hat{N}_w \rangle_- / N_{\rm e}$.  Given
these contributions, the total change of energy due to a single
feedback operation is then
\begin{equation}
  \label{eq:dE}
  \Delta E = 
  \left( \frac{\sigma^2}{2mN_{\rm e}} \!+\! 
    \frac{mN_{\rm e} \omega^2_0}{8\sigma^2}
  \right) \! \frac{\langle \hat{N}_w \rangle_-}{N_{\rm e}}
  + \frac{\langle \delta\!\hat{N}_w \hat{P}_w^2
    \rangle_-}{2m N_{\rm e}^2}
  - \frac{\langle \hat{P}_w^2 \rangle_-}{2m N_{\rm e}} .
\end{equation}

To achieve optimal cooling $\Delta E$ should be as negative as
possible which is achieved by minimizing the heating terms. The
optimal measurement resolution, $\sigma_{\rm opt} \!=\! \Delta p_0
\sqrt{N_{\rm e}}$, minimizes the measurement-induced heating [first
term in Eq.~(\ref{eq:dE})] to $\frac{1}{3} E_0 \langle \hat{N}_w
\rangle_- / N_{\rm e}$, where $E_0$ is the ground-state energy of a
single atom~\cite{remark}.  Choosing this value, the squared
measurement resolution per atom equals the squared ground-state
momentum uncertainty of a single atom, $\Delta p_0$.  Since the number
of atoms in the feedback region changes with temperature, the estimate
$N_{\rm e}$ should have a temperature dependence.  Thus $\sigma_{\rm
  opt}$ should be adapted during the cooling process, if possible, to
provide maximum cooling.  For further optimisation the size and
location of the feedback region in the $xy$-plane (cf.
Fig.~\ref{fig:geometry}) is of major importance. Here we consider only
particular cases; a more detailed study of the optimization will be
presented elsewhere. In particular, we focus on temperatures near the
condensation point, to demonstrate that in principle Bose--Einstein
condensation can be reached with stochastic cooling.

\begin{figure}
  \begin{center}
    \includegraphics[width=0.42\textwidth]{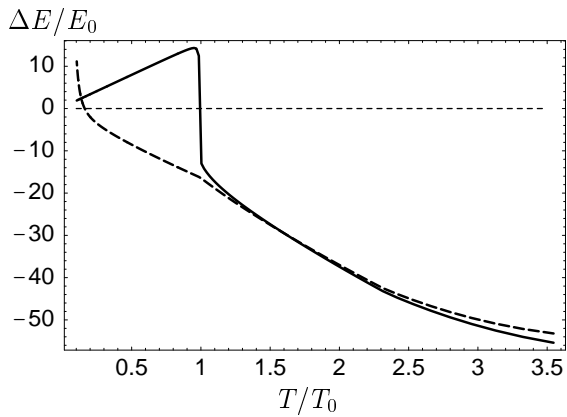}
    \caption{Energy change in units of $E_0$ versus
      temperature. $N_{\rm tot} \!=\! 10^6$, size of feedback region
      in $x$, $y$ directions is $\Delta q_0$, locations are: centered
      (solid), shifted out of center by $5 \!\times\! \Delta q_0$
      (dashed).}
    \label{fig:subtract}
  \end{center}
\end{figure}

We have calculated the expectation values in Eq.~(\ref{eq:dE}) using
the grand-canonical ensemble for a system at temperature $T$ and with
average total number of atoms $N_{\rm tot}$. In this way we obtain the
dependence of $\Delta E$ on the temperature for fixed $N_{\rm tot}$,
as shown in Fig.~\ref{fig:subtract} for $N_{\rm tot} \!=\! 10^6$
atoms.
If the feedback region is centered with respect to the trap potential,
it has substantial overlap with the condensate wavefunction. In this
case (solid curve), when the temperature is gradually decreased, there
is a sudden change from cooling to heating at the condensation
temperature $T_0$. This effect is due to dramatic atom-number
fluctuations of the condensate fraction at temperatures close to the
phase transition~\footnote{%
  See S.  Grossmann and M.  Holthaus, Phys.  Rev. E {\bf 54}, 3495
  (1996); M.  Wilkens and C.  Weiss, J.  Mod.  Opt.  {\bf 44}, 1801
  (1997).
%
}. These
fluctuations lead to a substantial heating due to the second term in
Eq.~(\ref{eq:dE}).

This problem can be avoided by choosing a feedback region that has no
spatial overlap with the nascent condensate and is thus not affected
by its large atom-number fluctuations. In this case the energy removed
per step gradually diminishes below $T_0$, though cooling still takes
place (dashed curve). On the other hand, the energy subtraction per
step at $T \!\gg\!  T_0$ is slightly smaller than for a centered
feedback region. An advantageous strategy is therefore to gradually
move the feedback region out of the trap center when approaching the
condensation temperature.

\begin{figure}
  \begin{center}
    \includegraphics[width=0.42\textwidth]{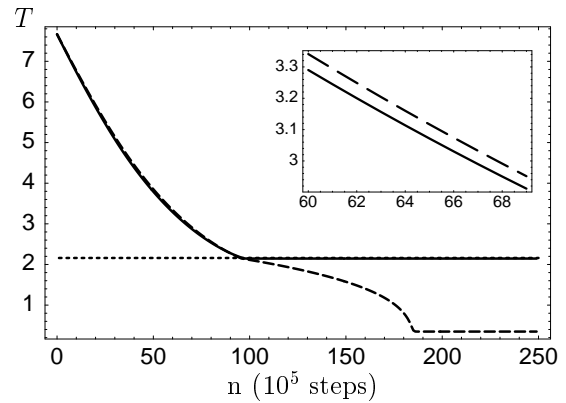}
    \caption{Temperature $T$ (in $\mu$K) versus number of feedback
      operations $n$. Parameters as in Fig.~2, $\omega_0/(2\pi) \!=\!
      500$Hz. Solid (dashed) curve corresponds to centered
      (non-centered) feedback region, doted line indicates $T_0$.}
    \label{fig:rate}
  \end{center}
\end{figure}

Our present results lead us to the conclusion that neither quantum
measurement effects nor the lack of knowledge of the precise number of
atoms in the feedback region in principle prevent Bose--Einstein
condensation by stochastic cooling. However, we have said nothing
about the speed at which condensation may be reached. A conclusive
answer on this issue requires that the cooling rate be dynamically
calculated using the method described in Ref.~\cite{wal-feedback}.
For now we proceed with an equilibrium approach that represents a
worst-case scenario: Starting at temperature $T_i$ we calculate the
energy subtraction $\Delta E(T_i)$, and thus a new average energy.
From this energy a new temperature $T_{i+1}$ is obtained by assuming
the equilibration of the system during the free evolution after the
feedback. That is, we assume that due to collisions the atoms exchange
energy and re-establish a new equilibrium state.
Iterating this calculation many times we obtain the dependence of
temperature on the number of feedback operations, as depicted in
Fig.~\ref{fig:rate}.
Starting with a temperature of $7.6 \, \mu$K, condensation is reached
after about $10^7$ feedback operations.  Again a feedback region
outside the trap center is advantageous when reaching $T_0$ (dashed
curve).

In a realistic experiment stochastic cooling is performed without
waiting for re-equilibration between feedback operations
\footnote{Remixing of atoms happens on the faster timescale
  $1/\omega_0$.}. Then non-vanishing coherent amplitudes of the
momentum $\langle \hat{P}_w \rangle$ will appear depending on the
oscillation phase, i.e., the (randomly chosen) time of the subsequent
feedback operation. Since $\langle \hat{P}_w^2 \rangle \!=\! \langle
(\Delta\hat{P}_w)^2 \rangle \!+\! \langle \hat{P}_w \rangle^2$, the
feedback then not only reduces momentum fluctuations $\langle
(\Delta\hat{P}_w)^2 \rangle$, as for a thermal state, but also
subtracts the energy due to the non-equilibrium coherent amplitudes
$\langle \hat{P}_w \rangle^2$ [cf.~Eq.~(\ref{eq:dE})]. The latter
leads to additional energy subtractions as compared to our equilibrium
approach. One can thus expect a much faster cooling process, so that
our results (cf.~Fig.~\ref{fig:rate}) represent the upper limit of the
required number of feedback operations.

Our calculations have shown that increasing the size of the feedback
region further increases the cooling efficiency.  Unfortunately, for
enlarged feedback regions an increased number of trap levels is
required, which presently runs into limitations of our numerics. For
finding optimal strategies and parameter ranges of stochastic cooling
we intend to implement in the near future a dynamical
calculation~\cite{wal-feedback}. Then the dependence of the cooling
rate on size and location of the feedback region can be studied in
full detail.

It is worth noting, that the considered feedback regions could be
easily realized in experiment by application of optical
fields~\cite{stochastic-raizen}. The ground-state position variance of
sodium atoms is approximately $\Delta q_0 \!=\!  0.6 \mu{\rm m}$ for a
trap of frequency $\nu \!=\! 500$Hz. Beam waists of externally applied
laser fields, that implement measurement and shift, can be chosen in a
wide range limited only by optical wavelengths, which are much smaller
than $\Delta q_0$.  Thus sizes of the feedback region much smaller and
much larger than $\Delta q_0$ could be realized.

Note also, that the geometry used here is different from that in
Ref.~\cite{stochastic-raizen}. Raizen {\it et al.} considered a
one-dimensional model where the feedback region was restricted in that
same direction.  They concluded that finer spatial resolution of the
feedback region leads to increasing cooling efficiencies. In our 3D
geometry that scenario would correspond to a restriction of the
feedback region also in $z$ direction, say to a size $\Delta z$. Then
the measurement can resolve momenta only within a resolution that is
enlarged by $1/(2 \Delta z)$. This however, may decrease the amount of
subtracted energy. The latter effect is absent in the classical
calculation as performed in Ref.~\cite{stochastic-raizen}. For now we
cannot confirm the predicted increase of cooling efficiency with
increased spatial resolution in $z$ direction, since such a scenario
would lead to non-trivial modifications in Eq.~(\ref{eq:dE}).

In conclusion by using a quantum-field theoretical approach we have
derived the energy change of feedback operations of stochastic cooling
of trapped bosonic atoms. Besides the heating due to the quantum
measurement and the sought subtraction of kinetic energy, we have
shown that stochastic cooling is strongly governed by a noise term
that is due to atom-number fluctuations in the feedback region.  This
effect becomes dominant at the condensation point where atom-number
fluctuations are large. This detrimental heating effect can be
ameliorated by the choice of the feedback region. It has been further
shown that condensation temperatures can in principle be reached. Our
results are based on an equilibrium approach and higher cooling
efficiencies are expected for a non-equilibrium dynamics as realizable
in experiment. Future investigations with fully dynamically solutions
will provide further insight into stochastic cooling with respect to
optimization of cooling rates.

This research was supported by Deutsche Forschungsgemeinschaft and
Deutscher Akademischer Austauschdienst.

\end{document}